\documentclass[aps,prl,twocolumn,superscriptaddress]{revtex4-1}

\usepackage{dcolumn}
\usepackage{amsmath,amssymb,amsthm,paralist}
\usepackage{graphicx}
\usepackage{appendix}
\usepackage{subfigure}
\usepackage{longtable}
\usepackage{url}
\usepackage{verbatim}

\begin{document}


\title{Mass Measurement of $^{56}$Sc Reveals a Small $A=56$ Odd-Even
Mass Staggering, Implying a Cooler Accreted Neutron Star Crust}


\author{Z.~Meisel}
\email[]{zmeisel@nd.edu}
\altaffiliation[Present address: ]{Department of Physics, University
of Notre Dame, Notre Dame, 46556 Indiana, USA}
\affiliation{National Superconducting Cyclotron Laboratory, Michigan State University, East Lansing, 48824 Michigan, USA}
\affiliation{Department of Physics and Astronomy, Michigan State University, East Lansing, 48824 Michigan, USA}
\affiliation{Joint Institute for Nuclear Astrophysics, Michigan State University, East Lansing, 48824 Michigan, USA}
\author{S.~George}
\affiliation{National Superconducting Cyclotron Laboratory, Michigan State University, East Lansing, 48824 Michigan, USA}
\affiliation{Joint Institute for Nuclear Astrophysics, Michigan State University, East Lansing, 48824 Michigan, USA}
\affiliation{Max-Planck-Institut f\"{u}r Kernphysik, 69117 Heidelberg, Germany}
\author{S.~Ahn}
\affiliation{National Superconducting Cyclotron Laboratory, Michigan
State University, East Lansing, 48824 Michigan, USA}
\affiliation{Joint Institute for Nuclear Astrophysics, Michigan State University, East Lansing, 48824 Michigan, USA}
\author{D.~Bazin}
\affiliation{National Superconducting Cyclotron Laboratory, Michigan State University, East Lansing, 48824 Michigan, USA}
\author{B.A.~Brown}
\affiliation{National Superconducting Cyclotron Laboratory, Michigan State University, East Lansing, 48824 Michigan, USA}
\affiliation{Department of Physics and Astronomy, Michigan State University, East Lansing, 48824 Michigan, USA}
\author{J.~Browne}
\affiliation{National Superconducting Cyclotron Laboratory, Michigan State University, East Lansing, 48824 Michigan, USA}
\affiliation{Department of Physics and Astronomy, Michigan State University, East Lansing, 48824 Michigan, USA}
\affiliation{Joint Institute for Nuclear Astrophysics, Michigan State University, East Lansing, 48824 Michigan, USA}
\author{J.F.~Carpino}
\affiliation{Department of Physics, Western Michigan University, Kalamazoo, 49008 Michigan, USA}
\author{H.~Chung}
\affiliation{Department of Physics, Western Michigan University, Kalamazoo, 49008 Michigan, USA}
\author{A.L.~Cole}
\affiliation{Physics Department, Kalamazoo College, Kalamazoo, 49006 Michigan, USA}
\author{R.H.~Cyburt}
\affiliation{National Superconducting Cyclotron Laboratory, Michigan State University, East Lansing, 48824 Michigan, USA}
\affiliation{Joint Institute for Nuclear Astrophysics, Michigan State University, East Lansing, 48824 Michigan, USA}
\author{A.~Estrad\'{e}}
\affiliation{School of Physics and Astronomy, The University of
Edinburgh, EH8 9YL Edinburgh, United Kingdom}
\author{M.~Famiano}
\affiliation{Department of Physics, Western Michigan University, Kalamazoo, 49008 Michigan, USA}
\author{A.~Gade}
\affiliation{National Superconducting Cyclotron Laboratory, Michigan State University, East Lansing, 48824 Michigan, USA}
\affiliation{Department of Physics and Astronomy, Michigan State University, East Lansing, 48824 Michigan, USA}
\author{C.~Langer}
\affiliation{National Superconducting Cyclotron Laboratory, Michigan State University, East Lansing, 48824 Michigan, USA}
\affiliation{Joint Institute for Nuclear Astrophysics, Michigan State University, East Lansing, 48824 Michigan, USA}
\author{M.~Mato\v{s}}
\altaffiliation[Present address: ]{Physics Division, International Atomic Energy Agency, 1400 Vienna, Austria}
\affiliation{Department of Physics and Astronomy, Louisiana State University, Baton Rouge, 70803 Louisiana, USA}
\author{W.~Mittig}
\affiliation{National Superconducting Cyclotron Laboratory, Michigan State University, East Lansing, 48824 Michigan, USA}
\affiliation{Department of Physics and Astronomy, Michigan State University, East Lansing, 48824 Michigan, USA}
\author{F.~Montes}
\affiliation{National Superconducting Cyclotron Laboratory, Michigan State University, East Lansing, 48824 Michigan, USA}
\affiliation{Joint Institute for Nuclear Astrophysics, Michigan State University, East Lansing, 48824 Michigan, USA}
\author{D.J.~Morrissey}
\affiliation{National Superconducting Cyclotron Laboratory, Michigan State University, East Lansing, 48824 Michigan, USA}
\affiliation{Department of Chemistry, Michigan State University, East Lansing, 48824 Michigan, USA}
\author{J.~Pereira}
\affiliation{National Superconducting Cyclotron Laboratory, Michigan State University, East Lansing, 48824 Michigan, USA}
\affiliation{Joint Institute for Nuclear Astrophysics, Michigan State University, East Lansing, 48824 Michigan, USA}
\author{H.~Schatz}
\affiliation{National Superconducting Cyclotron Laboratory, Michigan State University, East Lansing, 48824 Michigan, USA}
\affiliation{Department of Physics and Astronomy, Michigan State University, East Lansing, 48824 Michigan, USA}
\affiliation{Joint Institute for Nuclear Astrophysics, Michigan State University, East Lansing, 48824 Michigan, USA}
\author{J.~Schatz}
\affiliation{National Superconducting Cyclotron Laboratory, Michigan State University, East Lansing, 48824 Michigan, USA}
\author{M.~Scott}
\affiliation{National Superconducting Cyclotron Laboratory, Michigan State University, East Lansing, 48824 Michigan, USA}
\affiliation{Department of Physics and Astronomy, Michigan State University, East Lansing, 48824 Michigan, USA}
\author{D.~Shapira}
\affiliation{Oak Ridge National Laboratory, Oak Ridge, 37831 Tennessee, USA}
\author{K.~Smith}
\altaffiliation[Present address: ]{Department of Physics and
Astronomy, University of Tennessee, Knoxville, 37996 Tennessee, USA}
\affiliation{Joint Institute for Nuclear Astrophysics, Michigan State University, East Lansing, 48824 Michigan, USA}
\affiliation{Department of Physics, University of Notre Dame, Notre
Dame, 46556 Indiana, USA}
\author{J.~Stevens}
\affiliation{National Superconducting Cyclotron Laboratory, Michigan
State University, East Lansing, 48824 Michigan, USA}
\affiliation{Department of Physics and Astronomy, Michigan State University, East Lansing, 48824 Michigan, USA}
\affiliation{Joint Institute for Nuclear Astrophysics, Michigan State University, East Lansing, 48824 Michigan, USA}
\author{W.~Tan}
\affiliation{Department of Physics, University of Notre Dame, Notre
Dame, 46556 Indiana, USA}
\author{O.~Tarasov}
\affiliation{National Superconducting Cyclotron Laboratory, Michigan State University, East Lansing, 48824 Michigan, USA}
\author{S.~Towers}
\affiliation{Department of Physics, Western Michigan University, Kalamazoo, 49008 Michigan, USA}
\author{K.~Wimmer}
\altaffiliation[Present address: ]{Department of Physics, University of Tokyo, Hongo 7-3-1, Bunkyo-ku, 113-0033 Tokyo, Japan}
\affiliation{National Superconducting Cyclotron Laboratory, Michigan State University, East Lansing, 48824 Michigan, USA}
\author{J.R.~Winkelbauer}
\affiliation{National Superconducting Cyclotron Laboratory, Michigan State University, East Lansing, 48824 Michigan, USA}
\affiliation{Department of Physics and Astronomy, Michigan State University, East Lansing, 48824 Michigan, USA}
\author{J.~Yurkon}
\affiliation{National Superconducting Cyclotron Laboratory, Michigan State University, East Lansing, 48824 Michigan, USA}
\author{R.G.T.~Zegers}
\affiliation{National Superconducting Cyclotron Laboratory, Michigan
State University, East Lansing, 48824 Michigan, USA}
\affiliation{Department of Physics and Astronomy, Michigan State University, East Lansing, 48824 Michigan, USA}
\affiliation{Joint Institute for Nuclear Astrophysics, Michigan State University, East Lansing, 48824 Michigan, USA}


\date{\today}

\begin{abstract}


We present the mass excesses of $^{52-57}$Sc, obtained from recent time-of-flight nuclear mass
measurements at the National Superconducting Cyclotron Laboratory at
Michigan State University. The masses of $^{56}$Sc and $^{57}$Sc
were determined for the first time with atomic mass excesses of
$-24.85(59)(^{+0}_{-54})$~MeV and $-21.0(1.3)$~MeV, respectively, where the asymmetric uncertainty
for $^{56}$Sc was
included due to possible contamination from a long-lived isomer. 
The $^{56}$Sc mass indicates a small
odd-even mass staggering in the $A=56$ mass-chain towards the neutron drip line, significantly
deviating from trends predicted by the global FRDM mass model and
favoring trends predicted by the UNEDF0 and UNEDF1 density
functional calculations. Together
with new shell-model calculations of the electron-capture strength
function of $^{56}$Sc,
our results strongly reduce uncertainties in model calculations of the 
heating and cooling at the $^{56}$Ti electron-capture layer in the outer crust 
of accreting neutron stars. We find that, in contrast to previous studies,
neither strong neutrino cooling nor strong heating occurs in this
layer. We conclude that Urca cooling in the outer crusts of accreting
neutron stars that exhibit superbursts or high temperature steady-state burning,
which are predicted to be
rich in $A\approx56$ nuclei, is considerably weaker than predicted.
Urca cooling must instead be dominated by electron
capture on the small amounts of adjacent odd-$A$ nuclei contained in
the superburst and high temperature steady-state burning ashes.
This may explain the absence of strong crust Urca cooling inferred
from the observed cooling light curve of the transiently accreting
x-ray source MAXI J0556-332.
\end{abstract}


\maketitle


The thermal structure of the crust of neutron stars that accreted
matter from a nearby companion star directly relates to a number of 
astronomical observables, including the ignition of frequently
observed type-I x-ray bursts~\cite{Woos76,Grin76,Lamb78,Scha06,Pari13},
x-ray superbursts~\cite{Corn00,Cumm01,Stro02,Cumm04,Keek11,Keek12}, 
the observed cooling of transiently accreting neutron stars while accretion is 
turned
off~\cite{Brow98,Rutl99,Cack08,Cack10,Frid10,Page13,Dege14,Turl15}, and, potentially,  gravitational wave emission~\cite{Bild98,Usho00}.

The crust of accreting neutron stars strongly differs in composition
and thermal structure from isolated 
neutron stars. The composition is set by the ashes of hydrogen and helium 
burning on the surface via the rapid proton capture process (rp-process),
the $\alpha$p-process, and helium fusion
reactions~\cite{Scha99,Stev14}. With increasing depth, the 
rising electron Fermi energy $E_{\rm{Fermi}}$ induces electron-capture reactions  at specific locations where $E_{\rm{Fermi}}$ matches
the energy thresholds for electron capture. The result is a layered composition of more 
and more neutron rich nuclei that preserves the mass numbers $A$ of the thermonuclear ashes at the surface~\cite{Sato79,Blae90,Haen90,Gupt07,Stei12}. 
At still greater depths, beyond neutron-drip density, release and capture 
of neutrons, as well as pycnonuclear fusion reactions, lead to further 
changes in composition. While matter is accreted, these reactions operate 
continuously throughout the crust, maintaining its steady-state composition profile. The 
associated nuclear energy release heats the crust to higher temperatures 
than the neutron-star core. Alternatively, in some cases, depending on the nuclear
physics~\cite{Scha13}, an electron capture--$\beta$-decay Urca cycle~\cite{Gamo41} can occur in the thin
layer around a compositional boundary that leads to rapid neutrino
cooling instead of heating. 

Of particular importance
are the reaction sequences along the $A=56$ mass number chain in the outer crust. 
$A=56$ nuclei are predicted to make up a significant portion of the
outer crust in many neutron stars
because they are copiously produced for a range of hydrogen and helium burning 
conditions at the neutron star surface, including 
steady-state burning at high
accretion rates or high temperatures~\cite{Scha99} as is the case for the
quasi-persistent transient MAXI J0556-332~\cite{Homa14,Deib15}, type-I x-ray bursts,
and superbursts~\cite{Woos81,Scha99,Scha01,Scha03,Lang14}. 

\begin{figure}[h]
 \includegraphics[width=1.0\columnwidth,angle=0]{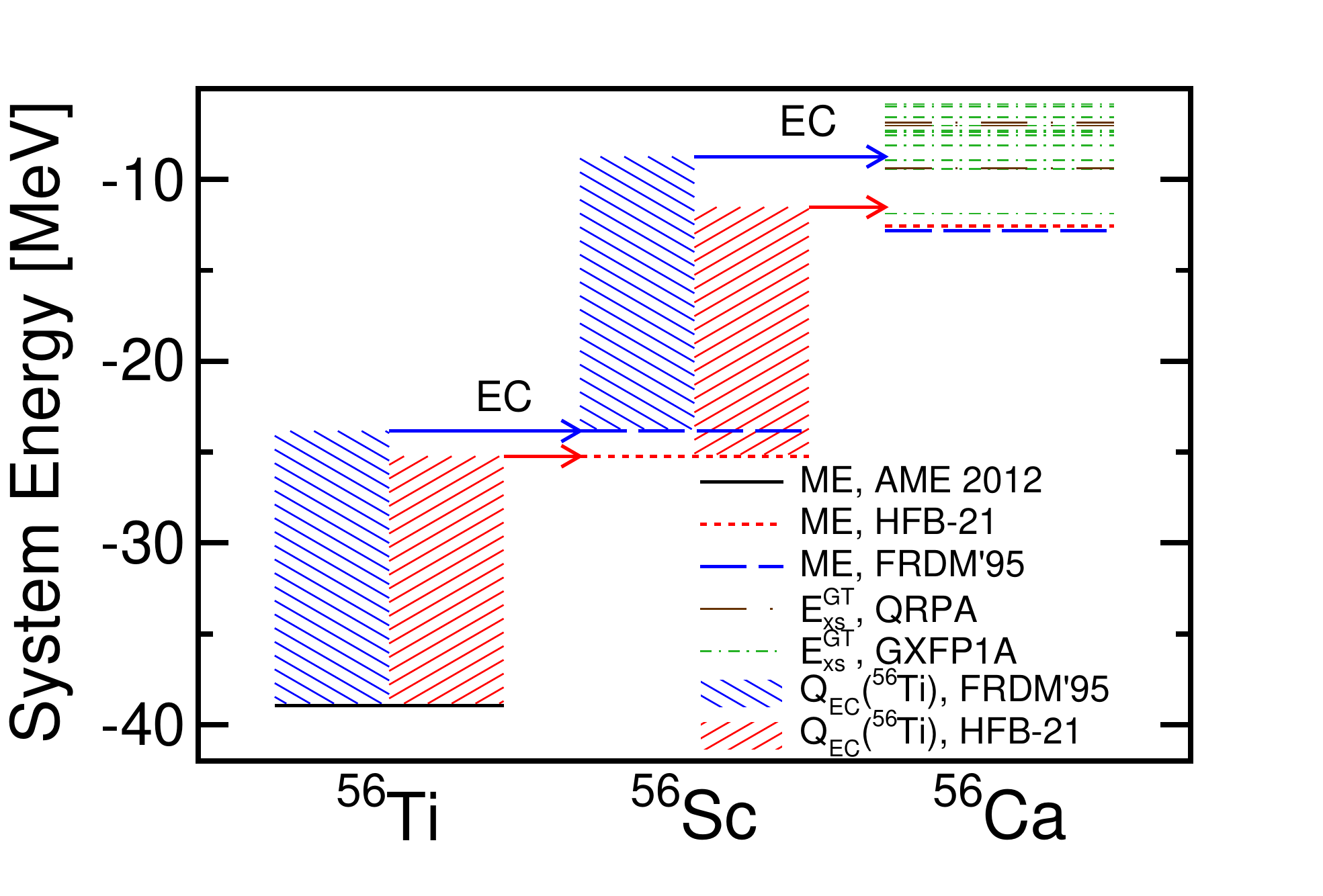}
 \caption{(color online). 
 Energy levels for the $A=56$ mass-chain at a depth where
 $E_{\rm{Fermi}}\approx |Q_{\rm{EC}}(^{56}$Ti)$|$, where atomic mass
 excesses $\rm{ME}$ are shown for the 2012 Atomic Mass
 Evaluation (AME)~\cite{Audi12} (solid black) if known experimentally and for
 theoretical mass models otherwise. 
 The larger odd-even mass staggering for the FRDM mass
 model~\cite{Moll95} (long-dashed blue) allows the second of the sequential electron
 captures (EC) to proceed through Gamow-Teller (GT) transitions,
 shown here for shell-model calculations using the GXPF1A
 Hamiltonian~\cite{Honm05} (dot-dashed green) and for the QRPA
 calculations used in~\cite{Scha13} (dot-dash brown), to
 higher-lying excited states $E_{\rm{xs}}^{\rm{GT}}$ in $^{56}$Ca than for the
 HFB-21 mass model~\cite{Gori10} (short-dashed red). 
 \label{EnergyLevelDiagram}}
 \end{figure}

\begin{figure}[]
 \includegraphics[width=1.0\columnwidth,angle=0]{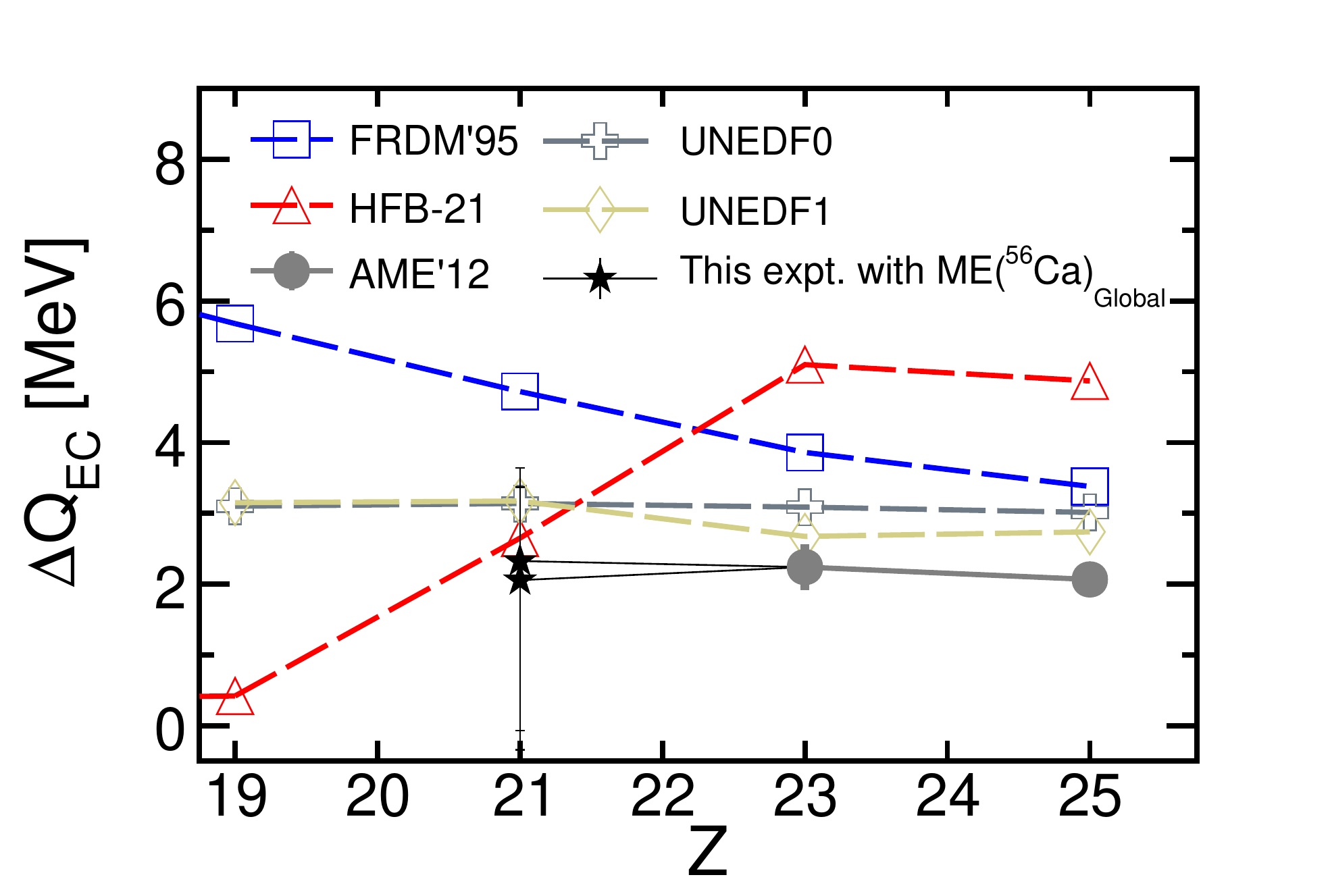}
 \caption{(color online). $\Delta Q_{\rm{EC}}(Z,A)$ for odd-odd $A=56$
 nuclei using ME($^{56}$Sc) from this experiment and ME($^{56}$Ca)
 from FRDM'95 or HFB-21 (black stars), compared to global
 mass models~\cite{Moll95,Gori10} and mass-differences predicted from recent energy density
 functional calculations~\cite{Kort10,Kort12} (open shapes). 
 \label{QECstaggerA56}}
 \end{figure}

As in all even-$A$ mass-chains, the odd-even staggering of electron-capture energy thresholds, 
a consequence of the nuclear pairing energy, leads to significant crust heating in the 
$A=56$ reaction chain. At a depth where $E_{\rm{Fermi}}$ just exceeds the threshold for 
electron capture $Q_{\rm EC}(Z,A)=\rm{ME}(Z,A)-\rm{ME}(Z-1,A)$ on an
even-even nucleus, the odd-odd nucleus formed by electron capture is immediately 
destroyed by a second electron-capture reaction with a lower threshold (see Fig.~\ref{EnergyLevelDiagram}). 
For this second step, 
$E_{\rm{Fermi}}$ exceeds the threshold, and the energy difference is split between 
the escaping neutrino and heat deposition into the crust. The energy release therefore 
corresponds directly to the magnitude of the odd-even staggering of
the electron-capture 
thresholds. For the $A=56$ chain of electron captures, thresholds are only known 
experimentally to $^{56}$Ti. Predictions for the odd-even staggering $\Delta Q_{\rm EC}=Q_{\rm EC}(Z-1,A)-Q_{\rm EC}(Z,A)$ beyond 
$^{56}$Ti vary dramatically (see Fig.~\ref{QECstaggerA56}). While
density-functional calculations 
predict a rather constant evolution of $\Delta Q_{\rm EC}$, the
FRDM~\cite{Moll95} mass model predicts a 
significant increase, and the HFB-21~\cite{Gori10} mass model predicts a dramatic drop. It is worth 
noting that even though $\Delta Q_{\rm EC}$ is a double difference of masses, predictions vary by 
almost 6~MeV, an order of magnitude larger than the sometimes quoted
global mass-model error~\cite{Moll95} and the related RMS deviations
of global mass model predictions from known masses.

Electron-capture thresholds are modified when the first available transition
proceeds through an excited state of the daughter nucleus, typically the lowest lying state with 
non-negligible transition strength~\cite{Gupt07}, rather than
through the ground state. In most cases, this does not change the general picture of 
a two-step electron capture sequence in even mass chains. However,
with the relatively small 
$\Delta Q_{\rm EC}$ between $^{56}$Ti and $^{56}$Sc predicted by the HFB-21 mass model
(2.65~MeV), and the relatively high 
excitation energy of the lowest-lying strength of the
$^{56}$Sc$\rightarrow^{56}$Ca transition (3.4~MeV) 
predicted by their global QRPA model, Schatz {\it et al.}~\cite{Scha13} point out an unusual situation 
where the electron capture on odd-odd $^{56}$Sc is blocked at the depth where 
$^{56}$Sc is produced by electron capture on $^{56}$Ti, preventing a
two-step electron 
capture sequence (see Fig.~\ref{EnergyLevelDiagram}). As a consequence, 
they~\cite{Scha13} find that $^{56}$Sc $\beta$-decay leads to a strong Urca cycle between 
$^{56}$Ti+$e^{-}\rightarrow ^{56}$Sc$ + \nu_e$ and
$^{56}$Sc$\rightarrow ^{56}$Ti$+e^{-}+\bar{\nu}_e$,
resulting in rapid neutrino cooling in neutron star crusts with $A=56$ material. This 
effect disappears when employing the large $\Delta Q_{\rm EC}$ predicted by the FRDM mass model 
\cite{Scha13}. 

To address the large uncertainties in the predicted 
$\Delta Q_{\rm EC}$ for $A=56$, we performed a measurement of the 
$^{56}$Sc mass. In addition, we carried out new shell-model calculations of 
the $^{56}$Sc electron-capture strength function which, in
connection with the new mass results,
lead to much improved predictions of heating and cooling of $A=56$ nuclei 
in neutron star crusts. 
 
 \begin{figure}[ht]
 \includegraphics[width=1.0\columnwidth,angle=0]{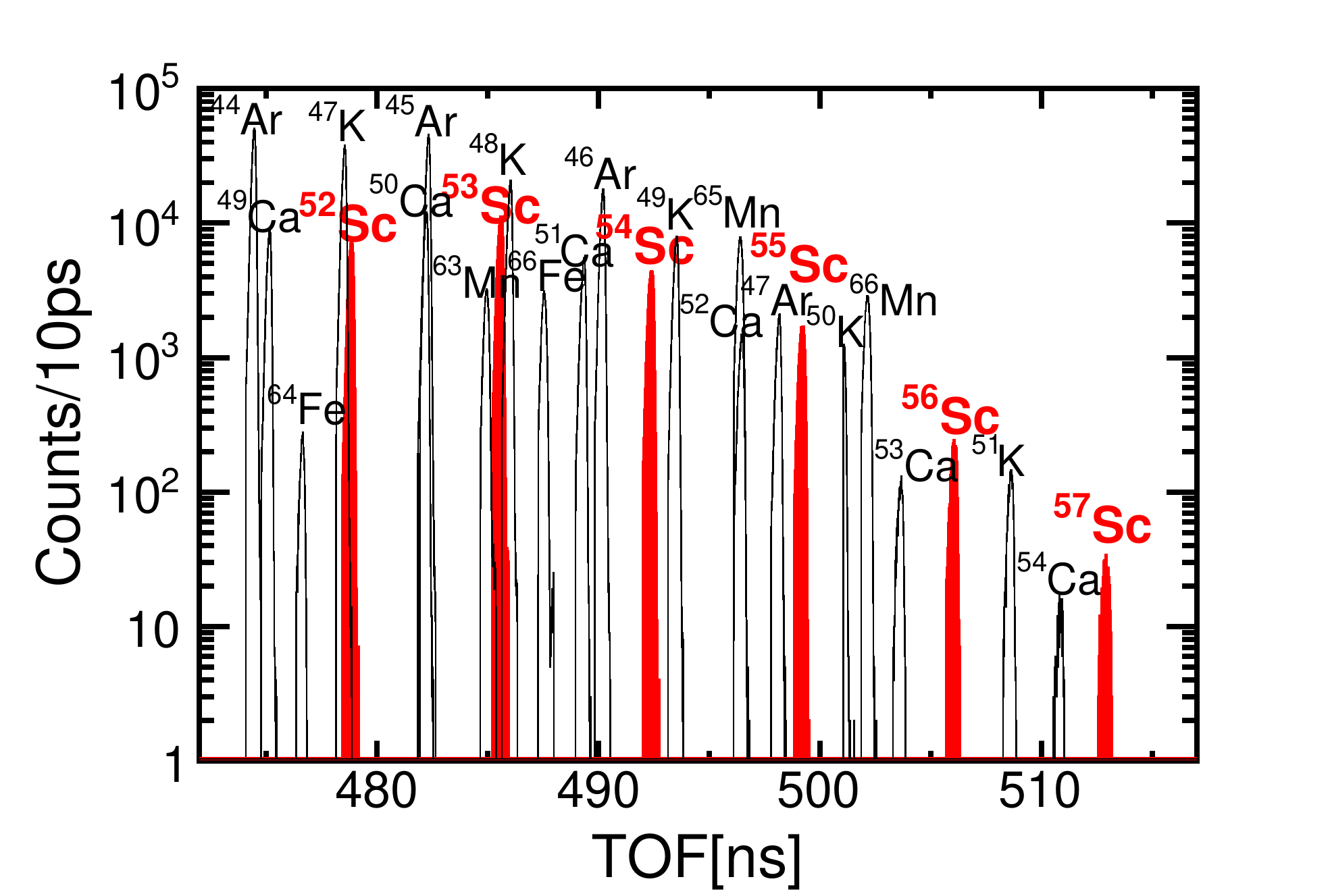}
 \caption{(color online). Rigidity-corrected time-of-flight distributions for
 reference nuclei (unfilled histograms) used to calibrate the
 $\frac{m_{\mathrm{rest}}}{q}(\mathrm{TOF})$
 relationship to obtain masses from the TOFs of the $A=52-57$
 isotopes of scandium (red-filled histograms).
 \label{TOFspectrum}}
 \end{figure}


 \begin{figure}[ht]
 \includegraphics[width=1.0\columnwidth,angle=0]{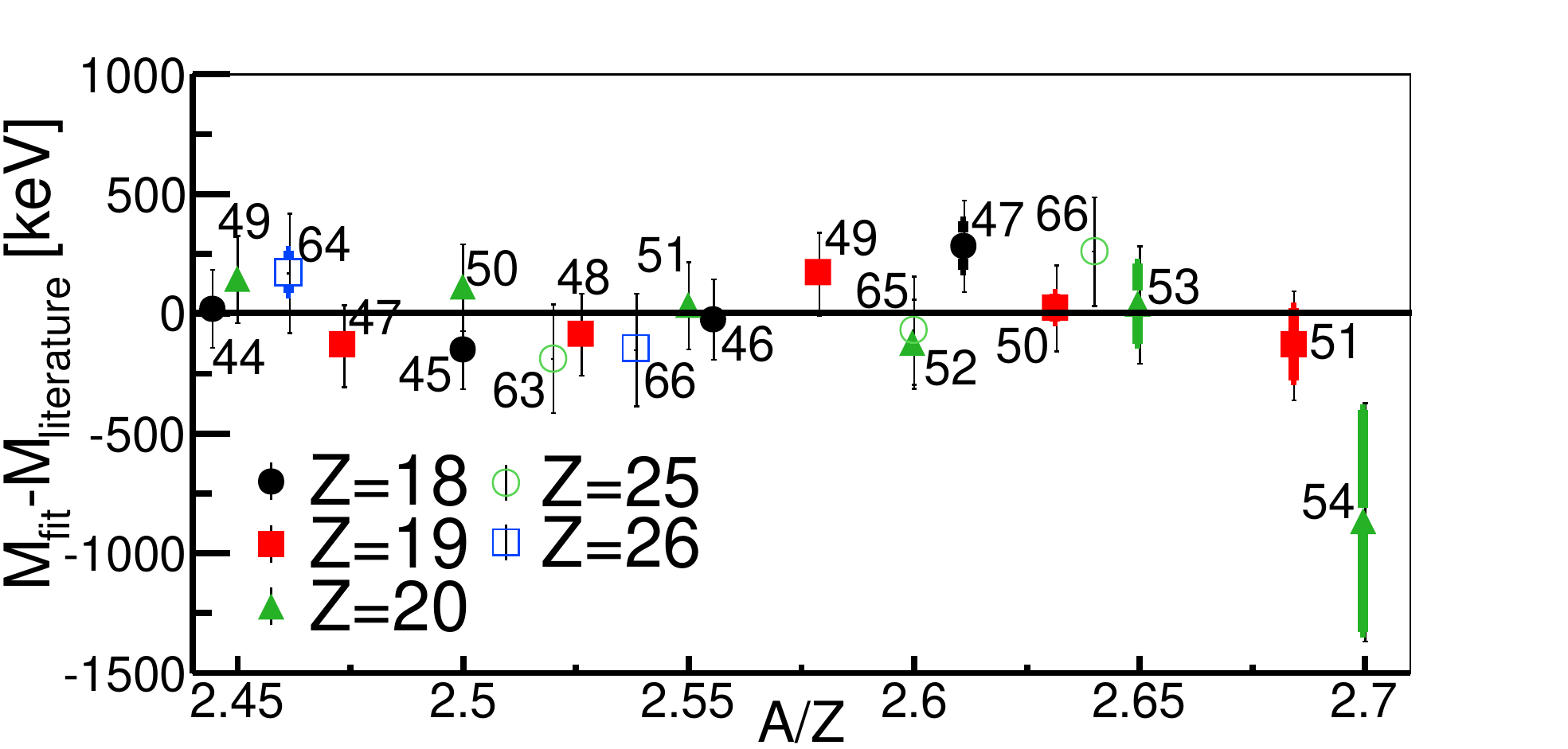}
 \caption{(color online). Residuals of the fit to the time-of-flight of calibration
 nuclei ($^{44-47}$Ar, $^{47-51}$K, $^{49-54}$Ca, $^{63,65,66}$Mn,
 and $^{64,66}$Fe) as a function of the
 mass number to nuclear charge ratio $A/Z$. Thick colored error bars show
 statistical uncertainties. Thin black error bars show the sum in
 quadrature of the statistical uncertainty and the systematic
 uncertainty, 9~keV/$q$ (here $q\equiv Z$), included for reference nuclei as described in~\cite{Mato12}.
 \label{MassFit}}
 \end{figure}

\begin{table*}[t]
  \caption{\label{ScMassComparison}
  Atomic mass excesses (in keV) of scandium isotopes measured in
  this experiment compared to results from previous direct mass
  measurements (TOFI~\cite{Tu90}, ESR~\cite{Mato04}, and
  NSCL~\cite{Estr11}), the adopted value in the
  2012 Atomic Mass
  Evaluation (AME)~\cite{Audi12} (`\#' are extrapolations), and predictions from global mass models
  (FRDM~\cite{Moll95} and HFB-21~\cite{Gori10}).
  The asymmetric uncertainty included for the $^{56}$Sc mass excess is an
  additional systematic uncertainty from potential isomeric
  contamination.}
  \def\arraystretch{1.25}
  \begin{ruledtabular}
  \begin{tabular}{lccccccc}
  Isotope & This expt. & TOFI & ESR & NSCL & AME 2012 & FRDM &
  HFB-21 \\ \hline
  $^{52}$Sc & $-40~300~(520)$ & $-40~520~(220)$  &       -       &
  -       & $-40~170~(140)$ & $-39~360$ & $-40~110$ \\
  $^{53}$Sc & $-38~170~(570)$ & $-38~600~(250)$  & $-38~840~(110)$ &
  $-38~110~(270)$ & $-38~110~(270)$ & $-36~840$ & $-38~480$ \\
  $^{54}$Sc & $-33~750~(630)$ & $-33~500~(500)$  & $-34~520~(210)$ &
  $-33~540~(360)$ & $-33~600~(360)$ & $-32~030$ & $-33~980$ \\
  $^{55}$Sc & $-30~520~(580)$ & $-28~500~(1000)$ &       -       &
  $-30~240~(600)$ & $-29~980~(460)$ & $-29~170$ & $-31~320$ \\
  $^{56}$Sc & $-24~850~(590)(^{+0}_{-540})$ &       -        &       -       &
  -       &$-24~731\#~(401\#)$& $-23~840$ & $-25~230$ \\
  $^{57}$Sc & $-21~000~(1300)$ &       -        &       -       &
  -       &$-20~707\#~(503\#)$& $-20~440$ & $-22~550$ \\
  \end{tabular}
  \end{ruledtabular}
\end{table*}

The masses of $^{52-57}$Sc were obtained with the time-of-flight
(TOF) method at the National Superconducting Cyclotron Laboratory~\cite{Mato12,Estr11,Meis13}.
The experimental setup and analysis are described 
in more detail in~\cite{Mato12,Meis13,Meis15} and are only summarized briefly here. 
A broad range of $\sim$150 neutron-rich isotopes from silicon to zinc
were produced by fragmentation of a 140~MeV/$u$ $^{82}$Se
beam on a beryllium target, transmitted through the 
A1900 fragment separator~\cite{Morr03}, and then sent to the focal
plane of the S800 spectrograph~\cite{Bazi03}. 
The fully-stripped ions were identified event-by-event using their time-of-flight (TOF) 
measured with fast-timing scintillators along a 60.6~m flight path
$L_{\mathrm{path}}$ and 
their energy loss in an ionization chamber. The magnetic rigidity
$B\rho$, being the ratio of momentum $p$ over charge $q$, of each ion was determined relative to
the tune of the beam line through a position 
measurement using a microchannel plate detector~\cite{Shap00} at the
dispersive focus at the S800 target position~\cite{Meis15}. 

The ion rest mass is related to TOF and $B\rho$  
through $m_{\mathrm{rest}}=\frac{\mathrm{TOF}}{L_{\mathrm{path}}}\frac{q(B\rho)}{\gamma}$, where
$\gamma$ is the Lorentz factor. Because neither $L_{\mathrm{path}}$ nor $B\rho$
are absolutely known with sufficient accuracy, the
$\frac{m_{\mathrm{rest}}}{q}(\mathrm{TOF})$ relationship is determined empirically
using reference nuclei with well-known masses~\cite{Meis13}.

The TOF distributions for reference nuclei and $^{52-57}$Sc are shown
in Fig.~\ref{TOFspectrum}.
Twenty reference nuclei with masses known
to better than 100~keV and no known isomeric states longer lived than
100~ns~\cite{Audi12,Wein13,Audi12B} were fitted with a 7-parameter calibration function of
first and second order in TOF, first order in TOF*$Z$, and containing first,
second, and fourth order $Z$ terms.
The calibration function represents a minimal set of terms that
minimized the overall fit residual
to literature masses and resulted in no detectable systematic
biases~\cite{Mato12}, as seen in Fig.~\ref{MassFit}. Additional
energy loss in the A1900 wedge degrader, which was not present
in~\cite{Mato12,Estr11}, required the addition of the TOF*$Z$ fit term. A systematic uncertainty
of 9.0~keV/$q$ was included as described in~\cite{Mato12} to normalize
the
$\chi^{2}$ per degree of freedom of the mass fit to one.
Two additional uncertainties related to the extrapolation were added
to the final mass uncertainties, one to reflect the uncertainties in the TOFs of
reference nuclei, which leads to an uncertainty in the fit
coefficients of the $\frac{m}{q}(\mathrm{TOF})$ relation, and one to reflect
the uncertainty inherent in
choosing a particular calibration function over another which has a
comparable goodness of fit.
The latter was determined by investigating the robustness of the
results to adding additional terms to the calibration function.
The total mass uncertainty is a sum in quadrature of
statistical, systematic, and two extrapolation uncertainties. The
relative contribution of the extrapolation uncertainties becomes
larger as the distance in $m/q$ and $Z$ from reference nuclei
increases.

The atomic mass excesses for scandium isotopes determined in this
experiment are compared to experimental and theoretical literature values in 
Tab.~\ref{ScMassComparison}. We note that the measured values
reported for $^{56}$Sc and $^{57}$Sc are a significant advancement over the
extrapolated values reported in the 2012 AME~\cite{Audi12}, as the
AME extrapolation assumes a locally smooth mass surface~\cite{Audi12}
and frequently fails in regions demonstrating changes in nuclear
structure (e.g. $^{53}$Ca and $^{54}$Ca~\cite{Wein13} as compared to the 2003
AME~\cite{Audi03}), such as the region covered by this work in which the $N=32$
and $N=34$ neutron sub-shell closures are weakly
constrained~\cite{Craw10}.
The mass uncertainties presented here correspond to a measurement precision of
$\delta m/m\approx 1\times10^{-5}$.
The primary contribution to the
overall measurement uncertainty comes from the uncertainty inherent
to the mass-fit extrapolation owing to the limited number of
reference nuclei with similar $Z$ and $A/Z$.
An additional uncertainty for ME($^{56}$Sc) originates from the
presence of a $\beta$-decaying isomer of unknown excitation energy~\cite{Lidd04,Craw10} that may be populated in the
fragmentation reaction producing $^{56}$Sc.  $^{56}$Sc has a
$\beta$-decaying low-spin (1$^{+}$) state and a $\beta$-decaying
high-spin ($5^{+}$ or $6^{+}$) state, but it is not known which is
the ground state and which is the isomeric state.  Shell-model
calculations with the GXPF1A Hamiltonian~\cite{Honm05} predict an excitation energy of the isomer of 540~keV.
The resolution of the $^{56}$Sc TOF peak is 100~ps, corresponding to
a mass resolution of 10~MeV, and can therefore not be used to
constrain the relative population of the ground and isomeric states.
Thus, the atomic mass excess obtained in this work represents a
least-bound limit for the $^{56}$Sc ground state and, guided by
theory, we add an
asymmetric uncertainty of $^{+0}_{-540}$~keV to our result to
account for the unknown population ratio.
The resulting atomic mass excess of $^{56}$Sc
determined in this work is $-24.85(59)(^{+0}_{-54})$~MeV.
As seen in
Tab.~\ref{ScMassComparison}, the atomic mass excess of $^{56}$Sc
presented here is consistent with the prediction from the
HFB-21~\cite{Gori10} global mass model, but is 
more bound than the prediction from the FRDM~\cite{Moll95} global
mass model.
 
Our result for $\rm{ME}(^{56}\rm{Sc})$ can be used to calculate the
odd-even staggering of $Q_{\rm{EC}}$ in the $A=56$ mass chain.
$Q_{\rm{EC}}(^{56}\rm{Ti})=-14.4(^{+1.3}_{-0.7})$~MeV is now determined
exclusively from experimental data. For $Q_{\rm{EC}}(^{56}\rm{Sc})$, we
still need the theoretical mass prediction for $^{56}$Ca. However, 
the large discrepancy for  $Q_{\rm{EC}}(^{56}\rm{Sc})$ between various mass
 models is exclusively due to the large discrepancies in the predictions for the 
 $^{56}$Sc mass, since predictions for the
atomic mass excess of $^{56}$Ca $\rm{ME}(^{56}\rm{Ca})$ agree within
$\approx$300~keV~\cite{Moll95,Gori10}. We therefore can 
combine our new $^{56}$Sc mass with the $^{56}$Ca mass predicted by
either the FRDM or HFB-21 mass models and find similar values of
$-12.0(^{+0.6}_{-1.1})$~MeV and $-12.3(^{+0.6}_{-1.1})$~MeV,
respectively. For the two choices of $^{56}$Ca mass, this results in
a $Q_{\rm{EC}}$ staggering of $\Delta
Q_{\rm{EC}}(^{56}\rm{Sc})=Q_{\rm{EC}}(^{56}\rm{Sc})-Q_{\rm{EC}}(^{56}\rm{Ti})=2.3(^{+1.3}_{-2.4})$~MeV
and $2.1(^{+1.3}_{-2.4})$~MeV, respectively.
Fig.~\ref{QECstaggerA56} shows the evolution of $\Delta Q_{\rm{EC}}$
in the $A=56$ mass chain for odd-$Z$ nuclei as a function of $Z$,
where we have included both of the aforementioned $\Delta
Q_{\rm{EC}}(^{56}\rm{Sc})$ in an attempt to capture the contribution
of the theoretical mass uncertainty of $^{56}$Ca. The new data
rule out the rapid increase in $\Delta Q_{\rm{EC}}$ approaching the
neutron drip line predicted by FRDM, and rather favor the predictions of recent energy
density-functional-based binding-energy
calculations~\cite{Kort10,Kort12} of a fairly constant $\Delta
Q_{\rm{EC}}$ along $A=56$.

\begin{figure}[]
 \includegraphics[width=1.0\columnwidth,angle=0]{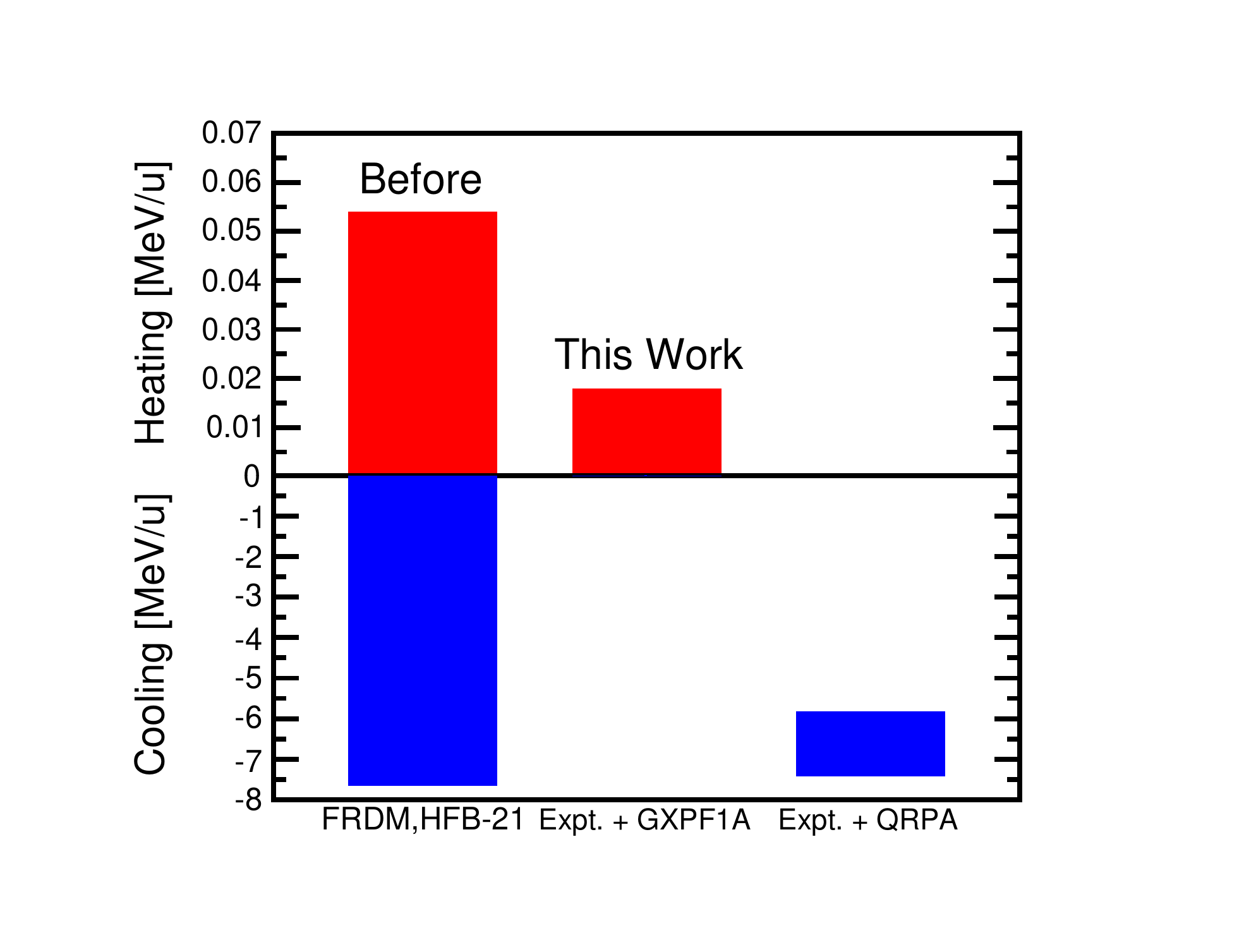}
 \caption{(color online). Integrated energy per accreted nucleon released from (negative
  values) or deposited into (positive
  values) the neutron star crust
  at the
  $^{56}\rm{Ti}\rightarrow^{56}\rm{Sc}\rightarrow^{56}\rm{Ca}$
  compositional transition 
  at a fiducial temperature of 0.5~GK and an accretion rate of
  0.3~$\dot{M}_{\rm{Eddington}}$.
  The left column indicates the large uncertainty prior to our work,
  where either heating or strong cooling were
  possible depending on the choice of global mass model for
  predicting $\rm{ME}(^{56}\rm{Sc})$ (FRDM~\cite{Moll95} or
  HFB-21~\cite{Gori10}) or on the choice of GT-transition
  strengths (shell-model using the GXPF1A
  Hamiltonian~\cite{Honm05} or the QRPA~\cite{Scha13}). The
  right and central columns show the narrow
  range of integrated heating/cooling possible when employing
  $\rm{ME}(^{56}\rm{Sc})$ reported here (within $\pm1\sigma$
  uncertainty)
  and GT-transitions from QRPA or the more reliable shell-model
  calculations performed for this study that employ the GXPF1A Hamiltonian.
 \label{integrated_heat_figure}}
 \end{figure}

The implications of $\Delta Q_{\rm{EC}}(^{56}\rm{Sc})$ obtained here
for the accreted neutron star crust were explored by inclusion of
our result for $\rm{ME}(^{56}\rm{Sc})$ in calculations performed
with the state-of-the-art crust
composition evolution model presented
in~\cite{Gupt07,Estr11,Scha13}. 
The model follows the compositional evolution of an accreted fluid
element with increasing pressure $p=\dot{M}\cdot g\cdot t$, where
the accretion rate $\dot{M}=2.64\times10^{4}$g/cm$^{2}$/s, surface
gravity $g=1.85\times10^{14}$cm/s$^{2}$, and time $t$, at a constant
temperature of $T=0.5$~GK (from~\cite{Gupt07}) using a full reaction
network that includes electron-capture, $\beta$-decay,
neutron-capture and their inverse, and fusion reactions.
These conditions are in the range inferred for the present
population of observed quasi-persistent transient
sources~\cite{Turl15}.
The $^{56}$Ti electron-capture layer was found to be either Urca cooling with more
than 7~MeV per accreted nucleon (HFB-21 mass model), or heating with
$0.05$~MeV per accreted 
nucleon (FRDM mass model)~\cite{Scha13} (see
Fig.~\ref{integrated_heat_figure}, `FRDM,HFB-21' column). The reason for this very large discrepancy is that 
in the FRDM mass model $\Delta Q_{\rm{EC}}(^{56}\rm{Sc})=4.3$~MeV is larger than the excitation 
energy of the lowest lying electron-capture transition in $^{56}$Ca predicted 
by the QRPA model used in previous studies (3.4~MeV), while in the 
HFB-21 mass model it is lower (2.6~MeV), as was demonstrated in
Fig.~\ref{EnergyLevelDiagram}. With the HFB-21 masses electron capture on 
$^{56}$Sc is therefore blocked initially and an effective Urca cycle 
involving $^{56}$Ti and $^{56}$Sc ensues. Our results for $\Delta
Q_{\rm{EC}}(^{56}\rm{Sc})$, when combined with the QRPA model, in principle are
closer to the HFB-21 case (see Fig.~\ref{integrated_heat_figure},
`Expt.+QRPA' column). 

However, heating and cooling at electron-capture transitions in neutron 
star crusts also depend sensitively on the electron capture and $\beta$-decay 
strength functions. In particular, the small odd-even mass stagger
for $A=56$ nuclei found in this work can lead to strong Urca cooling
		  depending on the location and strength of electron capture
		  and $\beta$-decay transitions. Previous studies employed predictions of a global QRPA model 
because of its availability for the entire range of nuclei of relevance for neutron 
star crusts. However, for the particular case of the electron capture on 
$^{56}$Sc of interest here, more reliable shell-model calculations
are possible~\cite{Cole12}. We 
performed such calculations using the GXPF1A effective
interaction~\cite{Honm05} and, using our new masses,
find no Urca cooling (see Fig.~\ref{integrated_heat_figure},
`Expt+GXPF1A' column). 
This is because the shell-model predicts a $1^{+}$ ground state for $^{56}$Sc and therefore
a strong allowed electron-capture transition to the ground state of $^{56}$Ca that removes
nuclei quickly from the $^{56}$Ti--$^{56}$Sc Urca cycle. 
Indeed, a $1^{+}$ ground state for $^{56}$Sc is consistent 
with experimental data, while the spin~3 prediction from the QRPA model is not. 

When using the shell-model strength function, our new $^{56}$Sc mass
significantly reduces uncertainties in predictions of nuclear
heating. In particular, it excludes the relatively strong heating 
predicted by the FRDM mass model, and limits heating to less than
$0.02$~MeV per accreted 
nucleon. Within mass uncertainties, no heating or even weak cooling
($0.002$~MeV/$u$) from pre-threshold
electron capture are possible. These results do not depend
significantly on our crust model assumptions, as heating is given
per accreted nucleon and is therefore independent of accretion rate,
and heating is relatively insensitive to the crust temperature.

In principle, experimental data do not exclude the possibility 
that the $1^+$ state in $^{56}$Sc is the long lived isomer and 
a 5$^+$ or 6$^+$ high spin state is the ground
state~\cite{Lidd04,Craw10}. In this case, selection rules would
prevent a ground-state to ground-state electron-capture transition
from $^{56}$Sc to $^{56}$Ca. However, even if the 
$1^{+}$ state in $^{56}$Sc is a low-lying, long-lived excited state
instead of the ground state, it will likely be thermally excited at temperatures in 
excess of 0.3~GK where 
Urca cooling is relevant, again leading to rapid depletion of the
$^{56}$Ti--$^{56}$Sc Urca cycle via an electron-capture transition
to the $^{56}$Ca ground state. Additionally, for the case of a
$5^{+}$ ground-state, the shell-model predicts a strong
electron-capture transition
into a 1.25~MeV excited state in $^{56}$Ca which could also be
populated by electron capture given our reported odd-even mass
stagger, thereby precluding Urca cooling as well.
Our shell-model based results are therefore robust. 

In summary, we have addressed the very large uncertainties in the
impact of the $^{56}$Ti electron-capture layer on the thermal
structure of accreting neutron star crusts reported in~\cite{Scha13} through a measurement of 
the $^{56}$Sc mass and shell-model calculations of the $^{56}$Sc
electron-capture 
strength. In contrast to previous studies, we find that neither strong cooling nor strong heating 
occurs in this layer. The thermal structure of accreting neutron
stars with superbursts or high temperature steady-state burning, which produce
large amounts of $A=56$ material, therefore depends sensitively on
the co-production of smaller amounts of odd-$A$ 
nuclei around $A=56$ that will dominate Urca cooling in the outer
crust.
To quantify this effect it is now crucial to reliably predict the abundance of odd-$A$ nuclei 
produced in the thermonuclear processes
on the surface of accreting neutron stars.

Overall, we find that Urca cooling in $A=56$-dominated accreted
neutron star crusts is much weaker than previously predicted. This
may explain the absence of strong Urca cooling recently inferred
from the x-ray cooling light curve of the transiently accreting
system MAXI J0556-332~\cite{Homa14,Deib15}, which is thought to host high temperature steady-state burning.

\begin{acknowledgments}
This project is funded by the NSF through Grants No. PHY-0822648,
PHY-1102511, PHY-1404442, and No. PHY-1430152.
S.G. acknowledges support from the DFG under Contracts No.
GE2183/1-1 and No. GE2183/2-1. We thank Erik Olsen for
providing nuclear binding energies from energy density
functional calculations and E.F.~Brown and A.T.~Deibel for many
useful discussions.
\end{acknowledgments}

\bibliographystyle{apsrev4-1}
\bibliography{TOFmassScUrcaMeisel}

\end{document}